# Distributed Metadata with the AMGA Metadata Catalog

Nuno Santos*
CERN
Nuno.Santos@cern.ch

Birger Koblitz†
CERN
Birger.Koblitz@cern.ch

## Abstract

*Catalog Services play a vital role on Data Grids by allowing users and applications to discover and locate the data needed. On large Data Grids, with hundreds of geographically distributed sites, centralized Catalog Services do not provide the required scalability, performance or fault-tolerance. In this article, we start by presenting and discussing the general requirements on Grid Catalogs of applications being developed by the EGEE user community. This provides the motivation for the second part of the article, where we present the replication and distribution mechanisms we have designed and implemented into the AMGA Metadata Catalog, which is part of the gLite software stack being developed for the EGEE project. Implementing these mechanisms in the catalog itself has the advantages of not requiring any special support from the relational database back-end, of being database independent, and of allowing tailoring the mechanisms to the specific requirements and characteristics of Metadata Catalogs.*

## 1. Introduction

File and Metadata Catalogs are essential services of a Data Grid, allowing users and applications to discover and locate data among the numerous sites of the Grid. File Catalogs map logical filenames to the physical location of one or more replicas of a file, while Metadata Catalogs store metadata describing the contents of the files, allowing users to search for files based on their description. Both of these services are essential for the operation of a Data Grid and, therefore, their dependability and scalability is of the greatest importance.

Providing dependable and scalable catalog services on a large Data Grid is challenging. For instance, the LCG [10] (LHC Computing Grid), one of the largest Data Grids in the world, consists of over two hundred computer centers distributed across the world. When it goes into full operation, thousands of jobs running concurrently will generate a very high load on the Catalogs. The network latency between grid sites can also limit the performance of the system, if users or jobs on a site must always query a remote catalog. Finally, in a distributed environment like a Grid, failures will be common, and catalog services will have to tolerate them. Under these conditions it is clear that a centralized system does not provide adequate service, and that replication and distribution mechanisms are needed. As Data Grids grow in size (number of users and amount of data) these features will become even more important.

We are studying these issues by designing and implementing replication mechanisms into the AMGA Metadata Catalog [12], which is part of the gLite [7] software stack of the EGEE [4] project. Instead of replicating the data in the underlying database back-end, we chose to build replication mechanisms into AMGA itself. This provides database independent replication and allows exploiting the specific features of metadata for optimizing the replication, by supporting features like partial replication and federation of catalogs. We use asynchronous, master-slave replication for better scalability in a geographically

*Partially funded by grant SFRH/BD/17276/2004 of the Portuguese Foundation for Science and Technology (FCT)
†Partially funded by the Bundesministerium fur Bildung und Forschung, Berlin, Germany

distributed environment.

The remainder of this paper is organized as follows. Section 2 briefly presents the AMGA Metadata Catalog. Section 3 discusses the motivation for our work, Section 4 describes the implementation of replication on AMGA. Section 5 discusses the related work on distributed Metadata Catalogs, and Section 6 concludes with an outlook of the ongoing work.

## 2. The AMGA Metadata Catalog

We will provide here only a brief overview of the AMGA Metadata Catalog, focusing on features relevant for the distribution mechanisms. Further details are available in [12].

AMGA began as an exploratory project to study the metadata requirements of the LHC experiments, and has since been deployed by several groups from different user communities, including High Energy Physics, Biomed and Earth Observation. More recently, AMGA was incorporated into the gLite software stack as the metadata catalog of the EGEE project.

A Metadata Catalog stores *entries* corresponding to the entity being described, typically files. These entries are described by user-definable *attributes*, which are key/value pairs with type information. Entries are not associated directly with attributes. Instead they are grouped into *schemas*, with the schemas holding the list of attributes that are shared by all their entries. AMGA structures metadata as an hierarchy, similar to a file-system. Directories play the role of schemas; they may contain both entries and other schemas. This hierarchical model has the advantages of being natural to users as it resembles a file-system, and of providing good scalability as metadata can be organized in sub-trees that can be queried independently. It is also the basis for partial replication, as will be explained later.

AMGA is a C++ multi-process server, with an extensible back-end that supports multiple database systems (currently PostgreSQL, Oracle, MySQL and SQLite). It offers two access protocols for clients: Web Services using SOAP and TCP streaming based on a text protocol similar to SMTP or TELNET.

## 3. Dependability and Replication in Data Grids Catalogs

The work presented here is motivated by the use cases we have identified while working with the EGEE user community. This is a large and active community, as shown by the number of applications being ported and deployed on the EGEE grid [3]. These applications represent how the Grid will be used in the next few years. They vary widely in requirements, complexity, size and maturity, but it is possible to define general trends. Grid Catalogs, both File and Metadata, are essential for most of them. Here, we will describe the requirements of the High Energy Physics (HEP) and Biomed communities, which are examples of two very different use cases. Most of the other use cases fall somewhere in between.

HEP applications use a large number of files, in the order of hundreds of millions, with metadata associated to them. In most cases, the metadata is read-only, with the number of writes (creation of new entries) being an order of magnitude lower than the number of reads. HEP users are spread geographically across over 200 sites, requiring special attention to deal with high-latency connections. Security is not a primary concern, as the metadata is not sensitive. Authentication is required to prevent denial of service attacks and for tracking users, but data is commonly sent as clear-text. For this class of applications, the main concern is scalability, performance and fault-tolerance. Writes can easily be performed in one or a few central catalogs, but reads are more frequent and must be offloaded to read-only replicas that are closer to the users. Partial replication is also an important feature, as replicating only the data needed by local users can often result in an order of magnitude decrease in the amount of replicated data.

Biomed [6] applications manage a much smaller amount of metadata, but have stronger requirements concerning security as their metadata often contains confidential information about patients. This metadata is generated in different geographical locations (hospitals or laboratories). Due to its sensitivity, it must be handled with extreme care. In particular, replicating the data either to a central catalog or to other replicas, would increase the exposure to attacks. A better solution is the federation of individual catalogs into a single virtual catalog, allowing data to remain secure at its site of origin, while providing transparent access to authorized users regardless of their location.

Figure 1 describes the main usage scenarios we have identified. *Full and partial replication* correspond to the HEP application case, where data is generated centrally and replicated for fault-tolerance and scalability. Partial replication is necessary for situations where remote users are only interested in part of the Metadata. This scenario can be implemented either by replicating part of the directory hierarchy, or by using a filter to specify an arbitrary subset of the data. The former takes advantage of the hierarchical structure of metadata in AMGA to replicate only the sub-trees required at the slave. Filtering is more generic and flexible, allowing the slave to specify arbitrary conditions that will be used by the master to select the logs shipped to the slave. In fact, partial replication of sub-trees is a special case of filtering, where the filter matches only a sub-tree.

*Federation* corresponds to the Biomed use case, where data generated in different Grid sites is federated as a single distributed catalog consisting of several physical catalogs. In this case, the remote nodes can be either physical or virtual replicas. In the first case the data is copied to the replica, while in the second no data is copied; instead the metadata commands executed on the slave are redirected to the master. Federation is described in more detail at the end of this section, after presenting the replication architecture of AMGA.

## 4. Architecture

Replication in AMGA follows asynchronous, master-slave model. Asynchronous replication is used for coping with the high latency of Wide-Area Networks, since synchronous replication is known [8] for its lack of scalability on WANs. Using asynchronous replication has the disadvantage that while updates are being propagated replicas may be inconsistent. This is not a significant problem for the applications we are considering.

Master-slave replication was selected because it is the simplest model that covers the needs of the majority of our target applications, which have simple write patterns. The master-slave model works well as long as writes are not common or originate from the same geographical location. We are also studying mechanisms for more efficient updates, as described in section 6.

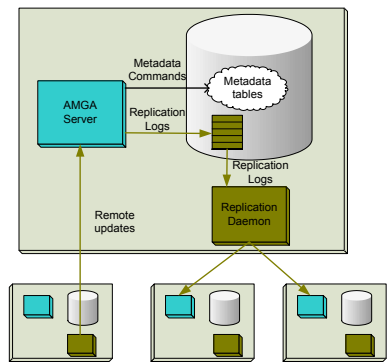

Figure 2: Replication internal architecture.

Figure 2 presents the replication architecture of AMGA. The basic idea is to have the master save on its local database a replication log for each metadata command that updates the back-end. These logs are then shipped to slaves that execute the command locally to bring themselves up-to-date. Since the metadata commands are independent of the database back-end, replication works even between AMGA servers using back-ends from different vendors. The AMGA Server is only responsible for saving replication logs. The remainder of the functionality is implemented on the replication module, which is an independent daemon that can run on a different machine for better performance.

**Managing subscribers** To replicate a directory from a master, a node must *subscribe* with the master to that directory. The slave can chose only a single directory or the metadata sub-tree rooted on a directory. Subscriptions are persistent; they outlive crashes of the master and of the slave. If a slave disconnects without having first requested to be unsubscribed, the master continues saving logs for the subscribed directories. When the slave reconnects, the subscription is resumed from the point it was interrupted. If the slave is disconnected for a long time, the master will eventually discard the subscription after either a configurable timeout or when the amount of pending logs ex-ceeds a certain threshold. If the master fails, the slave tries to reestablish the connection automatically, while continuing to serve its local clients. Currently there are no provisions for electing an alternate master during long outages, so the system is effectively in read-only mode during failures of the master.

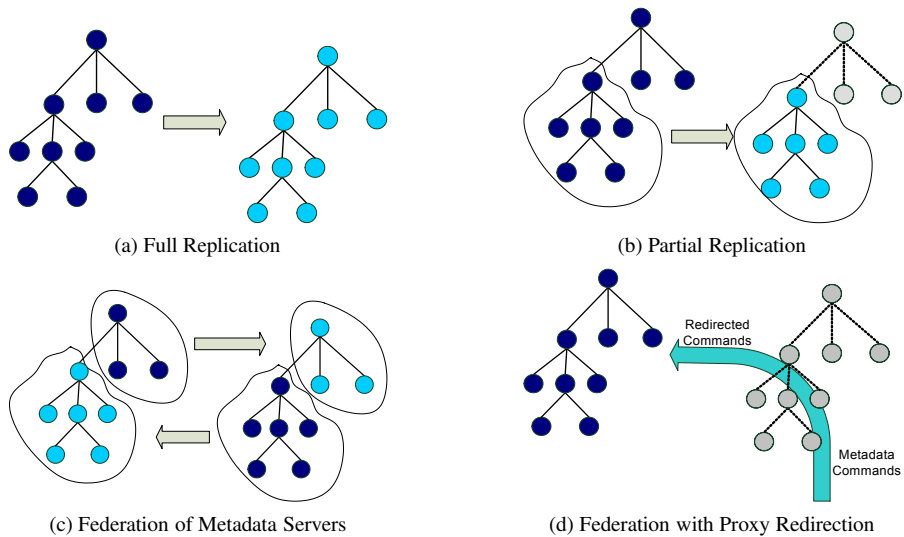

Figure 1: Replication/Distribution models.

**Generating logs** Each AMGA node having mastership of at least one directory writes the update logs to its database back-end. Each log entry is numbered with an unique sequence number and contains all the information required to execute the log on slaves. Logs are generated only for Metadata commands that result in updates to directories subscribed by some slave. To ensure consistency between the log table and the metadata tables, the log is written during the same database transaction used to perform the update.

**Sending the Initial Snapshot** After subscribing to a set of directories, the slave must copy from the master a snapshot of their contents. To remain database independent, we don't use the dump mechanisms offered by most databases, since they are database specific. Instead, we have implemented a similar dump command in AMGA that dumps the contents of a sub-tree of the metadata hierarchy as a sequence of metadata commands. While sending the snapshot the master continues accepting and executing requests from local users, which may result in updates to the directories that are being copied. To avoid inconsistencies the master uses a database transaction to isolate the sending of the snapshot from concurrent updates. During this transaction, it also reads the id of the last log generated and sends it to the slave, so that the slave knows where to start receiving logs. In the end, the slave will have a snapshot of the directory as it was when the transfer started, and will receive the updates performed in between as normal replication logs.

**Shipping Replication Logs** After obtaining the initial snapshot, the slave can start receiving and applying logs. In the current implementation the slave connects to the master using a TCP connection, sends the id of the first log it needs, and starts waiting. At the master, the replication module is responsible for shipping logs. It keeps track of all subscribers that are currently connected and of the id of the last log they have acknowledged. Periodically, it polls the log table and sends any new logs to the subscribers interested on them. The replication module also deletes the logs when they are no longer needed by any subscriber. When all subscribers are connected, logs are deleted shortly after being generated. Only when a subscriber is off-line will a log be kept for a longer time.

**Implementation of Federation** The mecha-

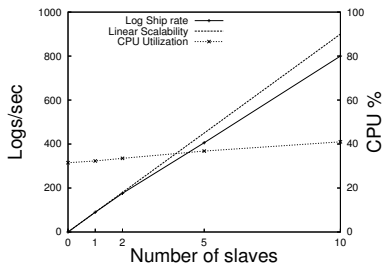

Figure 3: Scalability of a master serving up to 10 slaves. Insertion rate at master of 90 entries per second.

nisms described above support directly partial and full replication and provide the basis for federation. Federation exploits the hierarchical structure of metadata and the support for partial replication. Mastership is granted not to a catalog as a whole, but only to sub-trees of the catalog. Therefore, node A can be the master for directory /a, which is replicated by node B, while node B can be the master for directory /b, which can also be replicated by A. The distributed catalog contains both /a provided by A, and /b provided by B. These scheme permits different catalogs to have mastership of non-overlapping partitions of the metadata hierarchy, ensuring that for each directory there is a well identified master.

**Status and Performance Results** We have finished an initial implementation, which is currently available to the user community. Next are the results of a benchmark study assessing the scalability of the prototype. The benchmark was performed on a LAN. The master AMGA server plus the associated replication daemon were both running on the same computer, a P4 3GHz with 1GB of RAM. The slaves were running on two different computers, up to five in each, and were patched to discard incoming logs. This was done so that the slaves were not the bottleneck, since we wanted to test the scalability of the master. The tests were performed with the slaves already connected to the master and waiting for logs. We then inserted 10.000 entries on the master at a rate of 90 per second (corresponding to around 80% of the maximum rate when used standalone). We measured the time it takes for the logs to be sent to the slaves and the CPU load at the master. Figure 3 shows the results.

The data point for 0 slaves corresponds to having the master save replication logs for slaves that are subscribed but are disconnected. To provide a baseline, we measured the CPU usage with the AMGA server running standalone, i.e., no subscribers and not saving logs, which was around 20%. As can be seen, the scalability is close to linear with only a modest increase in CPU usage on the master.

## 5. Related Work

The Storage Resource Broker (SRB) [2] contains a metadata catalog service (MCAT) with support for federation and replication mechanisms [11]. The MCAT is an integral part of the SRB and cannot be easily reused on systems not based on the SRB software stack, while the AMGA catalog is designed to be a component in a modular architecture.

All major database systems have some kind of replication mechanism, like Oracle Streams or Slony-I for PostgreSQL, but they are all vendor specific and therefore do not address the heterogeneity of a Grid. The motivation for generating and shipping replication logs came from these systems. However, instead of replicating at the database level, we replicate metadata commands, which provides database independence.

The FroNtier [5] project aims to improve the performance of database read access over the Internet, by wrapping database queries in HTTP requests and using Internet caching mechanisms. The LCG's 3D project [9] is setting up a distributed database infrastructure for LHC experiments, using Oracle Streams as the main replication technology. Both Frontier and the 3D project are aimed at generic database applications while we are focusing on replication of Metadata Catalogs, which allows us to be database independent by moving the replication functionality from the database to the Metadata Catalog.

## 6. Conclusion and Future Work

We presented our work on replication and distribution mechanisms for Metadata Catalogs. These are essential to cope with the large number of users and the geographical distribution of grids, and for providing scalability and fault-tolerance. Although our work has focused on Metadata Catalogs, it is generic enough to be applied to File Catalogs. We have completed an initial prototype using the AMGA Metadata Catalog and have started testing with user communities.

We are studying other mechanisms for improving the dependability and scalability of the system. One clear limitation is the use of TCP connections between the slaves and the master for shipping logs. Group communication is better alternative for this situation, where a process needs to send messages to a potentially large number of nodes. It also offers better fault-tolerance, in the form of message delivery guarantees and retry mechanisms. After evaluating several messaging toolkits, we have decided to use Spread [1] and are currently working on updating our implementation.

The system must also detect and recover from node failures. If the failure is on a replica, clients should be redirected transparently to a different replica. If the failure is on the master, the remaining replicas should elect a new master among themselves after some grace period. All these mechanisms need an underlying discovery system so that replicas can locate and query each other, as well as mechanisms for running distributed algorithms among the nodes of the system.

We are also considering ways to support distributed updates. One possibility is multi-master replication using atomic broadcast to order updates. This solution is readily available as atomic broadcast is provided by group communication middleware, but might be expensive in terms of performance when done over Wide-Area Networks. The alternative is to extend the master-slave model to better support updates. If a slave receives an update request, it can either redirect the client to the master, contact the master and execute the update on behalf of the client, or acquire mastership and execute the update locally. We plan to study the behavior of these mechanisms under different usage scenarios.

## Acknowledgments

This work was performed within the LCG-ARDA project and the authors would like to thank Massimo Lamanna (CERN) for his valuable feedback and suggesting this interesting field of work in the first place. Viktor Pose (Dubna, Russia) performed intensive studies and testing of AMGA. We would also like to thank Peter Kunszt and Ricardo Rocha of the EGEE-gLite Data Management team for their collaboration in designing the Interface of AMGA, the GridPP team for its collaboration on metadata ideas. We received helpful comments about the paper from André Schiper and Craig Munro.

## References

[1] Y. Amir, C. Danilov, and J. Stanton. A low latency,loss tolerant architecture and protocol for wide area group communication. In *FTCS 2000*, 2000.

[2] C. Baru, R. Moore, A. Rajasekar, and M. Wan. The sdsc storage resource broker. In *CASCON '98: Conference of the Centre for Advanced Studies on Collaborative Research*. IBM Press, 1998.

[3] CERN. *EGEE Users Forum*, March 1-3 2006.

[4] EGEE - Enabling Grids for E-sciencE. http://public.eu-egee.org/.

[5] B. B. et al. Frontier: High performance database access using standard web components in a scalable multi-tier architecture. In *Computing in High Energy and Nuclear Physics*, Sept. 2004.

[6] J. M. et al. Bridging clinical information systems and grid middleware: a medical data manager. In *HealthGrid 2006*, June 2006.

[7] gLite - Lightweight Middleware for Grid Computing. http://glite.web.cern.ch.

[8] J. Gray, P. Helland, P. O'Neil, and D. Shasha. The dangers of replication and a solution. In *SIGMOD '96: Proceedings of the 1996 ACM SIGMOD international conference on Management of data*, pages 173–182, New York, NY, USA, 1996. ACM Press.

[9] LCG 3D Project - Distributed Deployment of Databases. http://lcg3d.cern.ch/.

[10] LHC Computing Grid (LCG). http://www.cern.ch/lcg.

[11] A. Rajasekar, M. Wan, R. Moore, and W. Schroeder. Data grid federation. In *PDPTA - Special Session on New Trends in Distributed Data Access*, pages 541–546, 2004.

[12] N. Santos and B. Koblitz. Metadata services on the grid. In *Advanced Computing and Analysis Techniques (ACAT'05)*, May 2005.